\documentstyle[12pt]{article}
\begin{document}
\baselineskip=24pt
\pagestyle{empty}
\begin{flushright}
RAL 92-073\vspace{10mm}
\end{flushright}
\begin{center}
\load{\Large}{\bf}
{\Large\bf
On the Validity of Perturbative Evolution of
\vspace{3mm} \\
Structure Functions from Low $Q^2$}
\vspace{10mm} \\
{\large J.R.Forshaw}
\vspace{10mm} \\
{\it Particle Physics Department,\\ Rutherford Appleton Laboratory,\\
Chilton, Didcot, Oxon,\\England.}
\vspace{10mm} \\
{\bf Abstract} \vspace{-3mm}
\end{center}
The validity of using QCD perturbation theory to generate dynamically the
parton distribution functions of hadrons, starting from a valencelike input at
low $Q^2$, is discussed. In particular, we consider the prescription of Barone
{\em et al} who evolve from $Q^2 = 0$, and that of Gl\"uck {\em et al} who
start
evolution from $Q^2 \simeq (2\Lambda_{QCD})^2$.
\vspace{5mm}
\begin{flushright}
November 1992
\end{flushright}
\newpage
\pagestyle{plain}
\pagenumbering{arabic}
\section{Introduction}
Recently, there has been significant interest in deriving parton (quark and
gluon) distribution functions of hadrons by dynamically evolving from
very low resolution scales \cite{BGNPZ,GRV}. The basic idea is to utilize the
fact that, at low resolution, hadrons appear to be a collection of valence
quarks. The details of the QCD dynamics allow one to generate the gluon and
sea components which are known to be present at higher resolution scales. Such
a program appears attractive since the input is reasonably well defined, and
much of the work is entrusted to perturbative QCD (pQCD). Compare this with
conventional approaches where one does not appeal to the valencelike structure
of hadrons at low resolution and is therefore left with the task of
constructing an input to the QCD evolution which must be extracted from the
data, e.g. see refs.\cite{trad}.

In this note, we wish to emphasise that great care must be taken when using
pQCD evolution from low $Q^2$ low resolution scales, and that previous attempts
are seriously flawed. In any perturbative calculation, one must be
sure to sum all of the relevant diagrams, and which class of diagrams is
relevant depends upon the kinematic regime under consideration.
Often, it is not sufficient to work to leading order in the coupling,
$\alpha_s$, because there may well be large logarithmic factors present which
seem to destroy the usefulness of $\alpha_s$ as an expansion parameter. The
need to sum an infinite subset of the perturbative expansion is quite commonly
encountered in pQCD calculations, in particular when calculating the dynamical
evolution of the distribution functions. We first will briefly review the
traditional calculation of the distribution functions, in particular for
deeply inelastic scattering (DIS) where the
spacelike virtuality of the photon $(-q^2=Q^2)$ provides the resolution scale.

In the parton model (where inter-parton correlations are negligible) the
factorisation of the DIS cross section into a hard (perturbative) piece and a
soft (non-perturbative) piece is straightforward -- Bjorken
scaling is predicted. As is well known, the violation of scaling is a
consequence of QCD corrections to the basic parton model. The naive
${\cal O}(\alpha_s)$ corrections to the basic parton model come from the
diagrams of fig.(1). However, a calculation of these diagrams reveals the
presence of logarithms $\sim \ln (Q^2/\mu^2)$ (for fixed $\alpha_s$),
where the scale $\mu^2$ is introduced to provide an
infra-red cutoff. For large $Q^2$, the presence of terms ${\cal O}(\alpha_s
\ln Q^2)$ seems to destroy the validity of a perturbative expansion.
Fortunately, we are able to sum up
the infinity of diagrams which possess a logarithm for each $\alpha_s$. In an
axial gauge, the contributors to this sum are the ladder diagrams, e.g. see
fig.(2). We are able to relate the distribution functions at
some scale $Q^2$ to their value at
another scale $Q_0^2$. Our ignorance regarding the soft physics is contained in
the input at $Q_0^2$. The choice of $Q_0^2$ must be sufficiently large to
ensure the validity of the subsequent evolution procedure. In the language of
the parton model, it is the Dokshitzer, Gribov, Lipatov, Altarelli and Parisi
(DGLAP) evolution equations which perform
this summation \cite{DGLAP}. In terms of the light-cone operator product
expansion (OPE), this summation is performed via the renormalisation group
equation, which relates the Wilson coefficients at different values of $Q^2$
(and hence the moments of the structure functions) \cite{RGE}.

One might attempt to start the pQCD evolution from some low resolution scale:
care must be taken. As one moves to lower scales, the presence of non-leading
logarithmic terms will be felt more and more, as will higher-twist terms.
Eventually, as $Q^2 \rightarrow \Lambda_{QCD}^2$, pQCD will breakdown as a
meaningful expansion. In the language of the OPE, the light-cone expansion
becomes less useful as $Q^2$ falls, since the dominant contribution is no
longer
on the light cone. In the next section, we concentrate on the parton
model picture of pQCD evolution and discuss how one expects the DGLAP equations
to fail at low $Q^2$. We discuss the modifications to DGLAP evolution
advocated by Barone, Genovese, Nikolaev, Predazzi and Zakharov (BGNPZ), who
claim to generate the parton content of hadrons
by evolving from $Q^2 = 0$ \cite{BGNPZ}. We conclude that significantly more
work is needed before one can claim to have even a reasonable phenomenological
model of evolution from $Q^2 = 0$. We also comment on the procedure of
Gl\"uck, Reya and Vogt (GRV), who evolve from
$Q_0^2 \simeq 0.3$ GeV$^2$ \cite{GRV}.

\section{QCD Evolution}
Let us show how the summation of leading logs is performed.
Consider the tree level process shown in fig.(1), where a quark from the parent
hadron radiates a real gluon. As is well known, one encounters singularities in
the cross section which must be regularised by taking into account the virtual
corrections of fig.(1). The final result is renormalisation
scheme dependent, it is leads to a modified quark distribution function
given by:
\begin{eqnarray}
q(x,Q^2) &=& q(x)+\frac{2\alpha_s}{3\pi} \int_{x}^{1} \frac{dy}{y} q(y) \left[
\left( \frac{1+z^2}{1-z} \right)_{+} \ln \frac{Q^2}{m_g^2} \right. \nonumber \\
&+& \left. (1+z^2)\left( \frac{\ln (1-z)}{1-z}\right)_{+} -2
\frac{1+z^2}{1-z}\ln z \right. \nonumber \\  &-& \left. \frac{3}{2(1-z)_{+}} +
4z+1-\left( \frac{2\pi^2}{3}+ \frac{3}{4} \right) \delta(1-z) \right].
\end{eqnarray}

The conventional `plus prescription' is used to describe the effect of the
virtual graphs and the non-logarithmic terms are determined in the massive
gluon regularisation scheme. The quark masses are neglected.

As the gluon mass vanishes, we have a logarithmic divergence. This can be
absorbed into a redefinition of the input, i.e. $q(y)\rightarrow q(y,\mu^2)$
where $\mu^2$ is some factorisation scale.
The perturbative expansion is only valid if
$Q^2$ is sufficiently large, i.e. it is usual to insist that $Q^2 \gg
\Lambda_{QCD}^2$. The presence of $\ln Q^2$ terms indicates that we should
treat all terms which are ${\cal O}((\alpha_s \ln Q^2)^n)$ on an equal footing.
They should be summed to ensure sensible results. Performing this summation,
and neglecting all those terms which do not lie within the LL approximation
leads to the DGLAP equations \cite{DGLAP}:
\begin{eqnarray}
\frac{\partial q_i(x,Q^2)}{\partial \ln Q^2} &=& \frac{\alpha_s(Q^2)}{2\pi}
\int_{x}^{1} \frac{dy}{y}\,(P_{qq}(x/y) q_i(y,Q^2) + P_{qg}(x/y) g(y,Q^2)), \\
\frac{\partial g(x,Q^2)}{\partial \ln Q^2} &=& \frac{\alpha_s(Q^2)}{2\pi}
\int_{x}^{1} \frac{dy}{y}\,(\sum_{j=1}^{2n_f} P_{gq}(x/y) q_j(y,Q^2) +
P_{gg}(x/y) g(y,Q^2)).
\end{eqnarray}
The splitting functions, $P_{ij}$, determine the probability for radiating a
parton of type $i$ from a parton of type $j$. For the process we considered,
the LL form for $P_{qq}$ is
\begin{equation}
P_{qq}(z) = \frac{4}{3} \left( \frac{1+z^2}{1-z} \right)_{+}.
\end{equation}
The strong ordering of transverse
momenta is inherent in these equations, and is the approximation which results
in selecting the $\ln Q^2$ terms which are essential for large $Q^2$, i.e.
\begin{equation}
k_{Ti}^2 \gg k_{Tj}^2
\end{equation}
is assumed. If one calculates the splitting functions to leading order (LO),
then one is selecting all terms which have one logarithm for each
$\alpha_s$, this is the leading logarithmic (LL) approximation. A
next-to-leading order (NLO) calculation of the splitting functions would
result in the inclusion of the next-to-leading logarithmic (NLL) terms, i.e.
those which are ${\cal O}(\alpha_s^n \ln^{n-1}Q^2)$. An example of a diagram
which contributes to the quark structure function in the NLL approximation is
shown in fig.(3).

It is clear that as $Q^2$ falls, the DGLAP equations run into serious
difficulties. BGNPZ attempt to modify the evolution, so that it remains finite
all the way down to $Q^2 = 0$. Let us outline their modifications. Note that we
do not simply reproduce their prescription, rather we present it what we
believe to be a more transparent way.

By appealing to the work of Gribov \cite{Gribov}, they do not permit the
coupling to become infinite as $Q^2 \rightarrow 0$. Rather, they introduce
some low momentum scale which causes the coupling to freeze at low $Q^2$, i.e.
they replace the leading order coupling with
\begin{equation}
\alpha_s(Q^2) = \frac{4\pi}{\beta_0 \ln((Q^2+k_0^2)/\Lambda_{QCD}^2)}.
\end{equation}
The scale $k_0^2$ is fixed by the requirement that it leads to the
experimentally observed pion-nucleon total cross section, i.e. $k_0 \approx
0.44$ GeV. In this case, $\alpha_s/\pi$ remains small enough that
perturbation theory may hopefully still apply.

The inclusion of quark masses is also necessary as $Q^2 \rightarrow 0$,
as is the inclusion of a gluon mass (which serves the purpose of regularizing
the gluon propagator, and confining the gluons). These are physical masses
which determine the scale $\mu^2$ in the $\ln (Q^2/\mu^2)$ factor. In this way,
they avoid pushing the physics below $\sim \Lambda_{QCD}$ into the
definition of the input.

To simplify things, it is assumed that one need only consider the radiation of
gluons from quarks, i.e. the splitting functions $P_{gg}$ and $P_{qg}$ are
neglected. This will be valid providing the gluon distribution function is
sufficiently small, which will be the case for not-too-small $x$.

Since partons which are radiated with very low transverse momenta occupy a
large transverse region of configuration space, it is possible that
interference terms, like the one in fig.(4) may become important. To this end
BGNPZ introduce a factor which is related to the two-quark form factor
of the valencelike hadron. This factor is very powerful in regularizing the
DGLAP kernel as $Q^2 \rightarrow 0$.

With the above modifications and simplifications in mind, the BGNPZ
prescription corresponds to using the
following evolution equations:
\begin{eqnarray}
\frac{\partial q_i(x,Q^2)}{\partial \ln Q^2} &=& \frac{\alpha_s(Q^2)}{2\pi}
\int_{x}^{1} \frac{dy}{y}\,\tilde{P}_{qq}(x/y)\, q_i(y,Q^2) \\
\frac{\partial g(x,Q^2)}{\partial \ln Q^2} &=& \frac{\alpha_s(Q^2)}{2\pi}
\int_{x}^{1} \frac{dy}{y}\,\sum_{j=1}^{n_f} \tilde{P}_{gq}(x/y)\, q_j(y,Q^2).
\end{eqnarray}
The freezing of $\alpha_s$ is understood to be operative and the modified
splitting functions are:
\begin{eqnarray}
\tilde{P}_{gq}(x) &=& V(x,Q^2)\frac{4Q^2}{3}
\frac{\{[1+(1-x)^2]Q^2+x^4m_q^2\}}{x[Q^2+(1-x)m_g^2+x^2m_q^2]^2},\\
\tilde{P}_{qq}(x) &=& \tilde{P}_{gq}(1-x).
\end{eqnarray}
The ggN vertex function is introduced to incorporate destructive interference
terms, i.e. long wavelength partons probe the colour singlet hadron and hence
decouple, it is given by
\begin{equation}
V(x,Q^2) = 1 - {\rm exp}\left(-\frac{R_{ch}^2}{2}
\frac{Q^2+x^2m_q^2}{1-x}\right),
\end{equation}
where $R_{ch}$ is the charge radius of the nucleon $(\sim 0.8$ fm).

Evolution is performed using the above equations starting from $Q^2 = 0$
assuming the nucleon to consist of three valence quarks only, i.e. their input
valence quark distribution is determined by the three-quark light-cone
wavefunction via
\begin{eqnarray}
q_i(x) &=& \int d^2{\bf k}_{t1} d^2{\bf k}_{t2} d^2{\bf k}_{t3} \,
\delta(\sum{\bf k}_{ti}) \nonumber \\
&\times& \int \frac{dx_2 dx_3}{x x_2 x_3} \delta(1-x-x_2-x_3) \left|
\Psi(x_i,{\bf k}_{Ti}^2) \right|.
\end{eqnarray}
They conclude that their results are relatively insensitive to the choice of
wavefunction, making both Gaussian and dipole ans\"atze. Clearly the attraction
of this approach is that the distribution functions
appear to be totally calculable in pQCD. The
inherent dependence upon the nucleon size is contained in the initial
wavefunction, and is the only non-perturbative parameter needed.

Of course, for high enough $Q^2$, one must regain the traditional DGLAP
equations. The $P_{qg}$ and $P_{gg}$ splitting functions are turned on at $Q^2
= 0.5$ GeV$^2$, where they expect the ggN vertex function to be close enough to
unity and neglect of the quark and gluon masses to be justified.

In the original paper, the QCD evolution is not presented in a way that is
quite so analogous to DGLAP evolution as the description above. Using the
above description of the BGNPZ model, it becomes evident that a number of
serious problems arise.

Inherent in the DGLAP approach, and the BGNPZ modification, is the
assumption of strong ordering in transverse momenta. There is no justification
in making this assumption if $Q^2$ is small, since the LL approximation is no
longer a good one. The evolution kernel should depend upon the transverse
momentum of the radiating parton, as well as on the radiated parton.

An example of an evolution equation which does not make the strong ordering
assumption is the Balitsky, Fadin, Kuraev and Lipatov (BFKL) equation, which
enables one to sum the diagrams relevant in the small $x$ domain
of QCD \cite{BFKL}. We emphasise that the construction of an
evolution equation necessitates that one is able to: (1) classify the set of
diagrams which need to be summed, and (2) derive those diagrams using basic
building blocks (which determine the evolution kernel). The BFKL equation is
designed to operate in the small-$x$ region, and the presence of large
logarithmic terms in $1/x$ (which can be classified) necessitates the
construction of an evolution equation which can be expected to sum the
dominant terms in the perturbative series. The BFKL equation has the structure:
\begin{equation}
\frac{\partial F(x,k^2)}{\partial \ln (1/x)} = \int dl^2 K(k^2,l^2)
F(x,l^2),
\end{equation}
where $F(x,k^2)$ must be integrated over $k^2$ to determine the gluon structure
function.

Away from small $x$, we expect the appropriate set of evolution equations to be
of the form:
\begin{equation}
F_{i}(x,k^2) = \int dl^2 dy\, K_{ij}(k^2,l^2,x,y) F_{j}(y,l^2).
\end{equation}
Since there are no large logarithms to sum we have no idea which set of
diagrams ought to be considered in deriving the kernel. The BGNPZ prescription
amounts to summing a rather arbitrary subset of diagrams, i.e. at low $Q^2$
there is no reason to single out those diagrams which are within the LL
approximation.

So, in the absence of any large logarithmic factors we are unable to single out
any particular subset of the perturbation series and have no real hope of
constructing a set of equations of the form determined in eqn.(14). To be
consistent therefore, we ought to use $\alpha_s$ as the expansion parameter.
The inclusion of the non-logarithmic terms (in eqn.(1) for example) is now
imperative, for they are no longer negligible relative to the $\ln (Q^2/\mu^2)$
term. Let us make this more explicit. Ignoring the factor $V(x,Q^2)$ (and the
running of $\alpha_s$), the BGNPZ prescription gives, for the quark
distribution function, logarithmic terms which are of the form
\begin{displaymath}
\ln \left( \frac{Q^2}{m_g^2} -1 \right)
\end{displaymath}
and
\begin{displaymath}
\ln \left( \frac{Q^2}{m_q^2} -1 \right)
\end{displaymath}
as the argument of the splitting function tends to zero and one respectively.
This is a direct consequence of assuming the strong ordering of momenta, i.e.
one integrates the quark virtuality over the range $0 < k_q^2 < Q^2$. The true
limits lead to a different logarithmic variation of the structure function, as
expressed in eqn.(1).

Thus for the BGNPZ prescription to make any sense one should abandon the strong
ordering assumption and keep all terms in the splitting function calculations,
using $\alpha_s$ as the expansion parameter. We no longer know how to derive
the evolution kernel. It should be recognised that there exist large logarithms
in $(1-x)$, which should be summed in order to ensure sensible behaviour as
$x\rightarrow 1$.

Compounding the problems further, since $\alpha_s$ is so large we expect (so
far
uncalculated) NLO contributions to be significant.
This point was realised in the slightly different case of LL and NLL evolution
by GRV \cite{GRV}. They emphasised the importance of considering
NLL corrections when evolving from $\alpha_s \simeq 0.9$.

All our discussions so far have been confined to leading-twist processes. There
are also higher-twist (HT) contributions (fig.(5)), which will depend upon the
multi-parton distribution functions.
There is no reason to neglect HT corrections at low $Q^2$, and it seems
reasonable to expect that their inclusion would lead to an enhancement of the
momentum carried by the $u$ quarks relative to the $d$ quarks within the
proton, (i.e. $uu$ pairs couple with spin-1, and $ud$ pairs with spin-0 or
spin-1, assuming a completely flavor symmetric quark distribution at some
scale, then higher-twist corrections result
in a lifting of the degeneracy of the spin-1 and spin-0 states
within the proton. The higher level is the spin-1 state and it
follows that the flavor symmetry is broken with $u$ quarks carrying a
larger fraction of the proton energy than one might naively expect \cite{FEC}).
Thus, even to first order in $\alpha_s$, the inclusion of HT terms seems a
necessary supplement to the BGNPZ approach.

We have so far emphasised the technical difficulties which one encounters when
attempting to evolve from low $Q^2$ (especially $Q^2 \sim 0$). There is also a
more fundamental difficulty, within the modified pQCD approach of BGNPZ,
which is concerned with the absence of any dynamical scale serving delineate
asymptotic freedom from confinement. As a clear example, consider the
following discussion.

In the case of the photon structure function, it is reasonably well
established by experiment that the photon (structure function) at low $Q^2$
resembles (that of) the $\rho^{0}$ (up to factors of $\alpha_{em})$ \cite{VMD}.
This leads to the
vector meson dominance hypothesis. Physically, one can understand such an
effect in terms of non-perturbative QCD. If the photon radiates a low-$p_T$
$q\bar{q}$ pair then gluon emission is favoured by the largeness of the
coupling $(\alpha_s(p_T^2))$ and the pair bind non-perturbatively to form a
vector meson. In the BGNPZ model, it is perfectly reasonable to emit a gluon
from a valence quark with a low $p_T$ (i.e. compared with the $p_T$ of the
$q\bar{q}$ pair discussed in the context of the photon). However, it is assumed
that no strong binding occurs subsequently between the gluon and valence quark,
which would appear to be in contradiction with the existence of a vector meson
contribution to the photon structure function.

The resolution of this paradox could be provided if one assumes that the
non-perturbative physics is added, by hand, at the outset. It is unlikely
that the BGNPZ modified perturbation theory, with non-perturbative
physics added independently is equivalent to traditional QCD, where the onset
of non-perturbative physics is signalled as the dynamical scale $Q^2$ tends to
$\Lambda_{QCD}$. We point out that the work of Gribov is intended to account
for confinement within a QCD-like framework -- it is not simply manifest by
freezing the coupling \cite{Gribov}.

To conclude, let us say a few words on the approach of GRV \cite{GRV}. Since
they start evolution at $Q^2 \simeq 4\Lambda_{QCD}^2$, the LL approximation may
well be useful. Indeed the dominance of the leading logarithmic terms is
supported by the NLL calculation, which (although seen to be significant)
results in a small correction to the LL result (for the structure function
$F_2$). However, the fact that the data seem to indicate the onset of
suppression due to the non-pertubative form factor
\begin{displaymath}
\left(\frac{Q^2}{Q^2+\nu^2}\right)^{\lambda}
\end{displaymath}
for $Q^2$ as high as 1 GeV$^2$ is worrying, and may well signal the importance
of HT effects below this $Q^2$. This should not be surprising, since a
conservative choice for $\nu^2$ would be $0.3$ GeV$^2$ and the Regge intercept
($\lambda$) is $1/2$ for valence quarks, giving a suppression factor of (at
least) 0.9 at $Q^2 = 1$ GeV$^2$, falling to (at least) 0.7 at $Q^2 = 0.3$
GeV$^2$.

It may well be that the GRV approach is unreasonable for $Q^2 < 1$ GeV$^2$
and is only designed to produce a structure function which fits the data at
$Q^2 \simeq 1$ GeV$^2$ (and hence beyond). If this is the case then one is left
with one of two conclusions. Firstly, it may be that, through a judicious
choice
of (valencelike) input, one is able to fit the high-$Q^2$ data more-or-less by
accident (if this is the case no benefit over more traditional structure
function analyses can be claimed). Secondly, given the clear importance of the
form factor suppression at low $Q^2$, one must conclude that the higher-twist
terms are effectively de-coupled from the leading-logarithmic leading-twist
terms, the origin of the de-coupling would then need to be explained.

Finally, although GRV claim to make serious small-$x$ predictions we feel this
to be wholly unjustified. The presence of large logs in $1/x$ cannot be ignored
in a perturbative analysis and one must therefore use the BFKL equation (with
appropriate shadowing corrections \cite{GLR}). The small $x$ regime of QCD is
a subject of much controversy, and we await the data which will soon come from
HERA to clarify the situation.

\section*{Acknowledgments}
I should like to thank Frank Close and Dick Roberts for useful discussions.

\newpage
\begin{flushleft}
{\bf FIGURE CAPTIONS}
\end{flushleft}
{\bf Figure 1} : Lowest order tree level amplitudes which contribute to the
quark-to-quark splitting function, and the virtual graphs which regularise the
$x\rightarrow 1$ singularities.
\vspace{5mm} \\
{\bf Figure 2} : A typical ladder graph, of the type that must be summed in
the leading log approximation.
\vspace{5mm} \\
{\bf Figure 3} : A typical contribution which must be considered in the
next-to-leading log approximation.
\vspace{5mm} \\
{\bf Figure 4} : Interference term between gluon distribution function
amplitudes. The gluons originate from different quarks.
\vspace{5mm} \\
{\bf Figure 5} : Higher-twist contribution, the calculation of which
necessitates an understanding of the diquark distribution function.
\end{document}